\documentclass{sig-alternate}
\usepackage{graphics}
\usepackage{epsfig}
\usepackage{subfigure}
\usepackage{latexsym}
\usepackage{amssymb}

\begin{document}

\title{Expressing OLAP operators with the TAX XML algebra}

\numberofauthors{3} 

\author{
\alignauthor Marouane Hachicha
\alignauthor Hadj Mahboubi\\
       \affaddr{Universit\'{e} de Lyon}\\
       \affaddr{(ERIC Lyon 2)}\\
       \affaddr{5 av. Pierre Mend\`{e}s-France}\\
       \affaddr{69676 Bron Cedex}\\     
       \affaddr{France}\\     
       \email{1st.last@univ-lyon2.fr}
\alignauthor J\'{e}r\^{o}me Darmont
}

\maketitle

\begin{abstract}

With the rise of XML as a standard for representing business data, XML data warehouses appear as suitable solutions for Web-based decision-support applications. In this context, it is necessary to allow OLAP analyses over XML data cubes (XOLAP). Thus, XQuery extensions are needed. To help define a formal framework and allow much-needed performance optimizations on analytical queries expressed in XQuery, having an algebra at one's disposal is desirable. However, XOLAP approaches and algebras from the literature still largely rely on the relational model and/or only feature a small number of OLAP operators. In opposition, we propose in this paper to express a broad set of OLAP operators with the TAX XML algebra. 

\end{abstract}



\section{Introduction}
\label{intro}

Data warehouses and OLAP (On-Line Analytical Processing) are nowadays technologically mature. However, their complexity makes them unattractive to many potential users, thus DSS (Decision Support System) vendors start developing simple, user-friendly Web-based interfaces. Furthermore, many decision-support applications (e.g., competitive monitoring) require data that are external to the institution exploiting them. In this context, the Web is a tremendous data source and a new trend toward on-line data warehousing is currently emerging, including approaches such as XML warehousing \cite{BoussaidMCA06,Pokorny02,ZhangWLZ05}.

The XML language is indeed becoming a standard for representing business data \cite{BeyerCCOPX05}. It is also particularly adapted for modeling \emph{complex data} from heterogeneous sources, and particularly the Web. Data we term complex are not only numerical or symbolic, but may be
represented in various formats (databases, texts, images, sounds, videos...);
diversely structured (relational databases, XML documents...);
originating from several different sources;
described through several channels or points of view (a video and a text that describe the same meteorological phenomenon, data expressed in different scales or languages...);
changing in terms of definition or value (temporal databases, periodical surveys...) \cite{DarmontBRA05}. Such data are much easier to logically model and store with XML than with relational models.


Many studies aim at performing OLAP analyses over XML data (XOLAP \cite{WangLHG05}). A first family mainly rely on the power of relational implementations of OLAP \cite{ChantolaAK07, JensenMP01, NiemiNNT02, PedersenPP04}. However, in these approaches, no or very few XML-specific OLAP operators are defined. Most recent studies directly relate to XOLAP \cite{ParkHS05, WangLHG05, WiwatwattanaJLS07}. However, each approach focuses on defining one cube operator only.
In this paper, we propose to express a broad set of OLAP operators with an existing XML algebra (namely, TAX \cite{JagadishLST01}), so that OLAP analyses can be applied onto native-XML data. On the long run, we are actually aiming at three objectives:
	(1) contribute to define a formal framework that does not currently exist in the XOLAP context;
	(2) support the effort for extending the XQuery language to allow OLAP queries, especially with XML-specific operators;
	(3) allow query optimization for OLAP XQueries. Native-XML DBMSs (Database Management Systems), though in constant progress, are indeed limited in term of performance and would greatly benefit from automatic query optimization, especially for costly analytical queries.

In this paper, we particularly focus on the first objective, i.e., defining XOLAP models. Computational issues shall be addressed later, but they are out of the scope of this paper, which is organized as follows. First, we present the context of this work and our motivation (Section~\ref{sec:TAX}). Second, we present the data model supporting our proposal (Section~\ref{sec:DataModel}). Then, we express OLAP operators in TAX and present three detailed examples of XOLAP operators (Section~\ref{sec:xolap-op-tax}). Finally, we compare our approach to X$^\wedge$3's \cite{WiwatwattanaJLS07}, which is in spirit the most closely related work to ours in the literature (Section~\ref{sec:comp-xolap-X3}). We conclude this paper and provide future research directions (Section~\ref{sec:ConclusionAndPerspectives}).



\begin{figure*}[t]
\centering
\epsfig{file=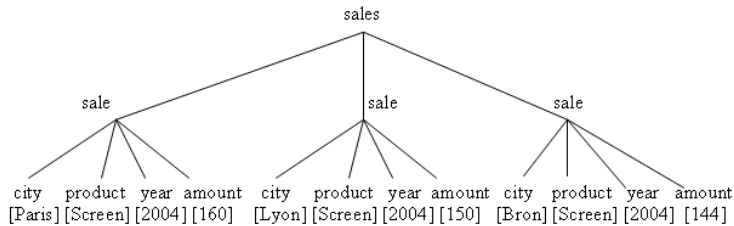, width=10cm} 
\caption{Sample TAX data tree}
\label{sample-tax-data-tree.eps}
\end{figure*}

\pagebreak

\section{CONTEXT and MOTIVATION}
\label{sec:TAX}

Rather than designing an XOLAP algebra from scratch, we chose to express a selection of ``basic" OLAP operators with an existing XML algebra. Many such algebras exist in the literature \cite{FernandezSW00,FrasincarHP02,NovakZ06}. Among them, we selected TAX (Tree Algebra for XML \cite{JagadishLST01}) for its richness. TAX indeed includes, under its logical and physical forms, more than twenty operators, which allows us many combinations for expressing XOLAP operators. Furthermore, TAX's expressivity is widely acknowledged. This algebra can be expressed with most XML query languages, and especially XQuery, which we particularly target because of its standard status. Finally, TAX and its derived algebra TLC \cite{PaparizosWLJ04} provide a query optimization framework that we can exploit in the future, since performance is one of our main concerns when desining decision-support applications that are integrally based on XML and XQuery. Unfortunately, due the space constraints, we cannot present the TAX elements that are necessary to understand our XOLAP operators in this paper. The interested reader may refer to the original paper describing TAX \cite{JagadishLST01} for full specifications.

In order to propose a broad range of XOLAP operators, we selected the most common OLAP operators to express them in TAX. When some of them had conflicting definitions, we selected the one that had the broadest acceptation. Before presenting them, we define in Section~\ref{sec:DataModel} the data model they are based on, which incorporates multidimensional concepts within the TAX data model.
The OLAP operators we selected to express in TAX are structural operators: rotate, switch, push and pull (Section~\ref{sec:StructuralOperators});
	set operators: slice and dice (Section~\ref{sec:SetOperators});
and	granularity-related operators: cube, roll-up and drill-down (Section~\ref{sec:GranularityRelatedOperators}). To express these operators, we adapted their formal definition in the classical OLAP context to the XML context by using the possibilities offered by TAX. Each XOLAP operator is actually represented as a combination of TAX operators. In this process, we took care of conserving the formal definitions and properties of OLAP operators, and used the TAX operators exactly as specified by TAX authors. 

We define each XOLAP operator formally and, for one of each type (namely, the rotate structural operator, the slice set operator, and the roll-up granularity-related operator), illustrate how it operates on the three levels of the data model: XOLAP cube (conceptual), XOLAP TAX tree (logical) and XML multidimensional document (physical). These multiple representations are aimed at empirically checking whether a given operator proceeds as expected. 
In Section~\ref{sec:xolap-op-tax}, we use the TAX notations introduced in \cite{JagadishLST01}. 

\section{Data model}
\label{sec:DataModel}

\begin{figure}[hbt]
\centering
\subfigure[XOLAP cube]{\epsfig{file=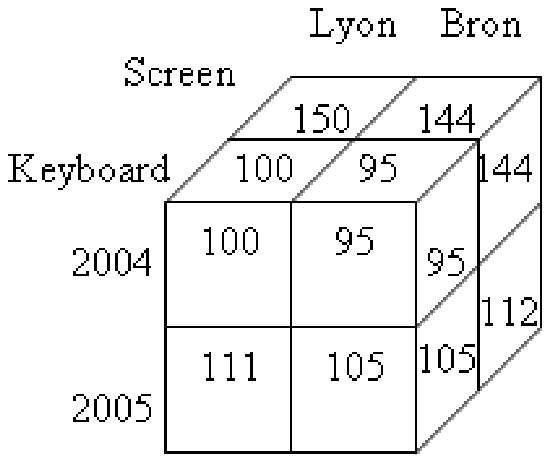, width=3cm}}\quad 
\subfigure[XML fact]{\epsfig{file=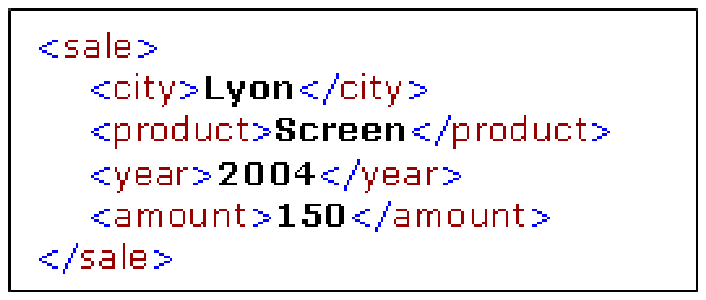, width=4cm}}\quad 
\caption{Sample XOLAP cube and XML fact}
\label{sample-cube-fact}
\end{figure}

A fact modeled in XML (or XML fact) is, classically, composed of dimension and measure elements. We term a set of XML facts an XML multidimensional document. Since TAX trees can model any XML document, XML multidimensional documents can be represented in TAX, in what we call multidimensional TAX trees, or XOLAP TAX trees. 
XOLAP TAX trees are thus a subset of TAX trees. An XOLAP TAX tree embeds a collection of fact elements, each described by dimension members and measure values. For instance, Figure~\ref{sample-tax-data-tree.eps} features an XOLAP TAX tree that contains \emph{sale} XML facts described by three dimensions: \emph{city} where the sale took place, sold \emph{product}, and sale \emph{year}. Selling \emph{amount} is the measure. Although logically modeled in XML, such facts are still conceptually multidimensional. Thus, we can represent an XML fact by a cube cell. We name such a cube an XOLAP cube, and its cells XOLAP cells. Figure~\ref{sample-cube-fact}(a) shows an example of XOLAP cube. The (physical) XML document corresponding to one of its XOLAP cells is also presented in Figure~\ref{sample-cube-fact}(b).

In classical OLAP, the concept of order of dimensions is very important. The switch operator's purpose is indeed to manipulate it. Dimensional order is easy to achieve in XML documents and tree representations. In our data model, we thus define a dimension reading direction for XOLAP TAX trees and XML facts: from left to right (in breadth first) and from top to bottom (in depth first), respectively. In addition, in both cases, fact measures must always be placed last in a subtree, with respect to these reading orders (e.g., amount in Figures~\ref{sample-tax-data-tree.eps} and \ref{sample-cube-fact}). This ordering has a very important role in the definition of some XOLAP operators, such as pull and push (Section~\ref{sec:StructuralOperators}).

\begin{figure}[hbt]
\centering
\epsfig{file=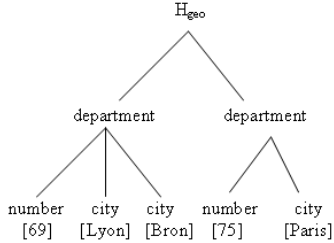, width=4.5cm} 
\caption{Sample hierarchical dimension specification}
\label{fig:H}
\end{figure}

Finally, our data model takes dimension hierarchies into account, to allow snowflake and constellation-like schemas. Each dimension's hierarchy is represented in a separate TAX tree. An example of such metadata TAX tree is provided in Figure~\ref{fig:H}. It refines the sample XOLAP TAX tree from Figure~\ref{sample-tax-data-tree.eps}'s geographical dimension by specifying that cities belong to departments (which are characterized in France by a department number). Note that, though the XOLAP TAX tree from Figure~\ref{sample-tax-data-tree.eps} represents only the finest granularity level (city), each city name is a reference to the geographical hierarchy TAX tree. We denote the set constituted of all dimension hierarchy trees $\cal H$.

\section{XOLAP OPERATORS IN TAX}
\label{sec:xolap-op-tax}

\subsection{Structural operators}
\label{sec:StructuralOperators}

\subsubsection{Rotate}
\label{sec:Rotate}

Rotate permutes the dimensions of a cube and simulates its rotation around one of its axes, so as to present different faces of the cube. In other words, rotating helps select cube faces rather than measures. In an XOLAP TAX  tree, we must exchange the positions of rotated dimensions for every subtree. Hence, the rotate operator may be expressed in TAX by a selection. Formally, its expression is: $Rotate(C) = \sigma_{P, SL}(C)$, where the rotated dimension is specified in pattern tree $P$ and dimension members and fact measures are specified in selection list $SL$.

\begin{figure*}[hbt]
\centering
\epsfig{file=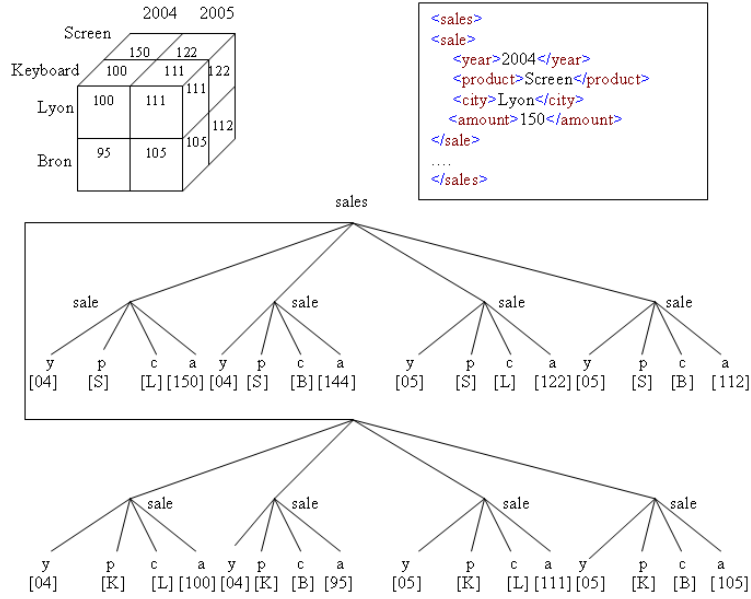, width=10cm} 
\caption{Rotate result example}
\label{rotate-example.eps}
\end{figure*}

Let us now take the example of rotating the XOLAP cube from Figure~\ref{sample-cube-fact}(a) around the \emph{product} dimension. Figure~\ref{rotate-example.eps} represents the rotated XOLAP cube (upper left), XOLAP TAX tree (bottom) and XML multidimensional document (upper right). In the selection, dimension order in the input pattern tree indicates the order of the edges in the output tree. Selection list $SL$ must contain the \textit{sale} node to retain all its children in output. 

\subsubsection{Switch}
\label{sec:Switch}

Switch exchanges the positions of two or several members of a given dimension. Like a rotation, this operation is mainly characterized by a visual effect and preserves measure values when changing the corresponding facts' positions. In an XOLAP TAX tree, we must reorder all subtrees. Hence, the switch operator may simply be expressed in TAX by using the reordering operator: $Switch(C) = \varrho_{P, o, RL}(C)$, where order TVF $o$ and reorder list $RL$ specify the new subtree order.

\subsubsection{Push}
\label{sec:Push}

Push associates the members of a dimension to measures of the cube, i.e., it transforms these members into cell contents. In TAX, push restructures the witness tree. For example, to transform the \textit{city} dimension into a measure, we associate the corresponding edge to the measure. Hence, the edge expressing this measure becomes composed of two subedges: \textit{city} and \textit{amount}. The push operator may thus be expressed in TAX with the copy-and-paste operator: \\$Push(C) = \kappa_{P, CL, US}(C)$, where copy list $CL$ and update specification $US$ specify how a dimension element is copied into the target fact element.

\subsubsection{Pull}
\label{sec:Pull}

Although pull is defined as push's reciprocal operator (transforming a measure into a dimension), it would be a mistake to express it in TAX push's reverse way. Moreover, in an XOLAP TAX tree, a measure is always represented by the last edge of the fact subtree (from left to right). Then, the pull operator just needs to move the measure before other dimensions. In TAX, this may be achieved with the projection operator, the new position being given by projection list $PL$ and pattern tree $P$: $Pull(C) = \pi_{P, PL}(C)$.

\subsection{Set operators}
\label{sec:SetOperators}

\subsubsection{Slice}
\label{sec:Slice}

A data cube may be seen as a set of slices arranged horizontally or vertically. The slice operator dissociates one or several of such slices with respect to one or several attributes from a given dimension. In TAX, we need to select the trees corresponding to these slices. As rotate, the slice operator may be expressed in TAX by the selection operator, but it is also combined to product to attach fact subtrees output by selection under the same root. Moreover, in the rotate operator, the selection operator's pattern tree $P$ indicates dimension order in the output, whereas in the slice operator, it determines the slicing dimension values. Formally, the expression of the slice operator is: $Slice(C) = \times(\sigma_{P, SL}(C))$. 

\begin{figure*}[hbt]
\centering
\epsfig{file=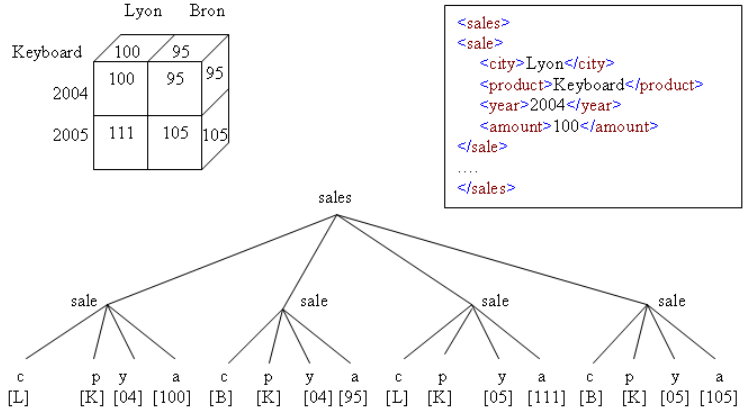, width=10cm} 
\caption{Slice result example}
\label{slice-example.eps}
\end{figure*}

As an example, let us extract, from an initial XOLAP tree, the subtrees corresponding to slices of the corresponding XOLAP cube. Figure~\ref{slice-example.eps} shows how the XOLAP cube from Figure~\ref{sample-cube-fact}(a) is sliced on the Keyboard instance of the \emph{product} dimension. It represents the sliced XOLAP cube (upper left), XOLAP TAX tree (bottom) and XML multidimensional document (upper right). Selection pattern tree $P$ determines the slicing dimension member value (Keyboard, here). Selection list $SL$ must contain the \textit{sale} node to retain all its children in output. Finally, the resulting \textit{sale} subtrees must be attached under a common root (\textit{sales}) with the product operator.

\subsubsection{Dice}
\label{sec:Dice}

The dice operator extracts a subcube from a data cube with respect to predicates defined on its dimensions. In other words, dicing extracts measures from a cube for a given number of members, which must be identical in all dimensions. In TAX, we may express the dice operator by selecting the corresponding subtrees from the XOLAP TAX tree with a simple pattern tree $P$ and attaching them to the same root with a product: $Dice(C) = \times(\sigma_{P, SL}(C))$. Note that this is again different from the slice operator, in which $P$ specifies the slicing member element(s). 


\pagebreak

\subsection{Granularity-related operators}
\label{sec:GranularityRelatedOperators}

\subsubsection{Roll-up}
\label{sec:roll-up}

Roll-up represents a given data cube at a higher (more general) level of granularity, with respect to a dimension hierarchy (e.g., moving from the city granularity to the department granularity along the geographical dimension from Figure~\ref{fig:H}). During this transition, measures must be aggregated with respect to a grouping operation. In a TAX OLAP tree, this translates as replacing all subtrees with the subtrees resulting from the roll-up operation. The roll-up operator may be expressed in TAX by a combination of the selection, grouping, join, aggregation, node deletion and node insertion operators. Formally, its expression is: $Roll\_up(C, H) = \iota_{P_\iota, IS}(\delta_{P_\delta, DS}(A_{P_A, g, US}(\Join_{P_\Join, SL_\Join}(\gamma_{P_\gamma, g, o}(\sigma_{P_\sigma, SL_\sigma}(C)), H))))$, where $P_\Join$ specifies the rolled-up dimension and the selected granularity level, and $H \in \cal H$ this dimension's hierarchy.

\begin{figure*}[hbt]
\centering
\epsfig{file=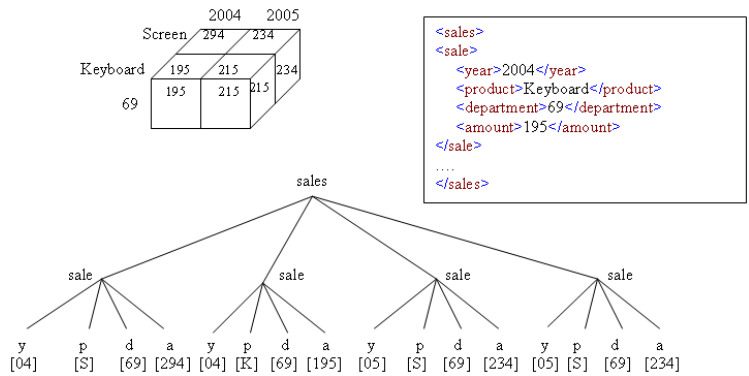, width=10cm} 
\caption{Roll-up result example}
\label{rollup-example.eps}
\end{figure*}

Its execution is illustrated by the following example: perform a roll-up on the \emph{department} dimension (department number 69, here) of the XOLAP TAX tree from Figure~\ref{sample-tax-data-tree.eps}. Thus, grouping occurs on (\textit{product}, \textit{year}). We use the \textit{sum} aggregation function on \emph{amount}. Figure~\ref{rollup-example.eps} represents the rolled-up XOLAP cube (upper left), XOLAP TAX tree (bottom) and XML multidimensional document (upper right). The corresponding sequence of TAX operator combination is detailed thereafter.


	(1) \emph{Selection:} We must first separate subtrees from the input to prepare them for the next step.
  (2) \emph{Grouping:} Then, we group the different subtrees by \textit{product} and \textit{year}. Grouping is achieved with respect to all possible combinations of modalities in both dimensions.  
	(3) \emph{Join:} For each tree resulting from grouping, we join subtrees corresponding to the rolled-up dimension (i.e., \emph{city}, here)
 and to the corresponding dimension hierachical tree $H$ (to retrieve department 69, here). 
	(4) \emph{Aggregation:} Now, for each tree resulting from the previous step, we aggregate all measures (i.e., \emph{amount} here) in one node by using the specified aggregation function (\textit{sum}, here).  
	(5) \emph{Tree updates:} Finally, we must prune from the output tree the measures at the previous granularity level, which are retained (node deletion) and add a new node describing the resulting granularity level (\emph{department} in our case) with node insertion. 


\subsubsection{Drill-down}
\label{sec:drill-down}

Drill-down is roll-up's reciprocal operator. It helps refine a cube dimension at a lower aggregation level, e.g., moving from the department granularity to the city granularity along the geographical dimension from Figure~\ref{fig:H}. First, the drilled-down dimension must be selected, then it must be joined to its hierarchy specification $H \in \cal H$ and the finest-granularity cube $C_0$ (to retrieve non-aggregate measure values). Then, we must use the projection operator to mask the remaining coarser level member. Finally, measures must be reaggregated to the correct hierarchy level and all subtrees are attached to a single root (product). Formally, the drill-down operator is expressed as follows: $Drill\_down(C, H) = \times(A_{P_A, g, US}(\pi_{P_\pi, PL}(\Join_{P_\Join, SL_\Join}(\sigma_{P_\sigma, SL_\sigma}(C), H, C_0))))$.

\subsubsection{Cube}
\label{sec:cube}

Cube is an aggregation operator that generalizes grouping, roll-up and join on a data cube to produce a more general (aggregate) cube. In TAX, we first must complete a roll-up for each cube dimension (Section \ref{sec:roll-up}). Then, resulting subtrees must be grouped with respect to every possible combination of dimensions. Finally, these subtrees are joined into one tree, with all measures being aggregated. The grouping, join and aggregation  TAX operatiors are combined as follows: $Cube(C) = A_{P_A, ag, US}(\Join_{P_\Join, SL_\Join}(A_{P_A, ag, US}(\gamma_{P_\gamma, g, o}(Roll\_up(C, \cal H)))))$.

\section{Comparison to X$^\wedge$3}
\label{sec:comp-xolap-X3}

Let us eventually discuss the differences between X$^\wedge$3 \cite{WiwatwattanaJLS07} and our own XOLAP approach, which have been developed in parallel without knowledge of the other effort. First, we use TAX to support our proposal by expressing each XOLAP operator with one TAX operator or more. On the other hand, Wiwatwattana \emph{et al.} exploit the principle of TAX's grouping operator, but X$^\wedge$3 stands as is and is not based on TAX operators. Thus, it addresses the issues of summarizability and heterogeneity in XML tree structures in an ad-hoc fashion, while our approach is more general.

Another difference between X$^\wedge$3 and our approach lies at data model level. While we exploit TAX's pattern and witness trees, Wiwatwattana \emph{et al.} have based X$^\wedge$3 on relaxed tree patterns, which were initially introduced for approximate XML query matching \cite{AmerYahiaCS02}. Hence, X$^\wedge$3 outputs a lattice whose nodes represent a cuboid, while our result trees are simpler. We indeed do not explicitely take the specificities of the XML model into account and rather trust TAX to do so instead.

Finally, to take on an algorithmic metaphor, X$^\wedge$3's authors have adopted a depth-first approach by fully developping one operator only (which is of little use alone), while we have adopted a breadth-first approach by proposing a wider range of operators (which only apply onto quite ``regular" XML data without ragged hierarchies nor missing values, though). Both approaches should aim at completion, in breadth and depth, respectively, to achieve a full XOLAP environment. We actually think they are quite complementary and should be combined, since our XOLAP approach provides modeling/logical functionalities, while X$^\wedge$3 bestows computational support on native XML data.

\section{Conclusion and perspectives}
\label{sec:ConclusionAndPerspectives}

In this paper, we have completed a first step toward an XOLAP framework, by initiating a previously inexistant formal background (Section~\ref{intro}, objective~1). To achieve this goal, we have demonstrated how the TAX XML algebra could support OLAP operators, by expressing in TAX the main usual OLAP operators (cube, rotate, switch, roll-up, drill-down, slice, dice, pull and push). By doing so, we significantly expanded the number of available XOLAP operators, since up to now, related papers only proposed at most three operators each (always including the cube operator).


The perspectives opened by this work are numerous. The main issue we are currently working on is to adapt the classical OLAP operators we expressed in TAX to actual specificities of XML. More precisely, we are addressing the four following issues, essentially when performing roll-up and drill-down operations:
	(1) ragged hierarchies \cite{BeyerCCOPX05} in dimensions, i.e., multivalued and multi-granularity dimension members;
	(2) facts that are defined at heterogeneous levels of granularity (e.g., sales might be detailed at the city level in some countries, while only available at the region level in others);
	(3) missing (unavailable) dimension information in facts;
	(4) facts with changing dimensional order (e.g., time and location in one fact, location and time in the next).
Our final objective here is to support the XQuery extension effort for decision-support applications (Section~\ref{intro}, objective~2). 

Though a non-trivial task that we have postponed for now, we also plan complexity analyses to provide a sounder comparison between the XOLAP operators from the literature and our own proposal. Furthermore, to allow experimentally validating our set of XOLAP operators, we have also started implementing it into a software prototype that helps generate the corresponding XQuery code. This simple Web-based querying interface is currently coupled to the TIMBER XML-native DBMS, but it is actually independent and could operate onto any other DBMS supporting XQuery. Our prototype currently features the rotate, slice and roll-up operators.

This software plaform shall also allow us to initiate a cycle of experimental cost evaluations and optimizations for each XOLAP operator, to find the most efficient expression of each XOLAP operator in TAX (there are, of course, often several solutions for expressing a given XOLAP operator in TAX). In our first experiments, the efficiency of XQuery indeed proved limited when processing complex analytical queries. Hence, we also plan on the long run to exploit our OLAP TAX expression as a basis for automatically optimizing XOLAP queries expressed with XQuery (Section~\ref{intro}, objective~3).

\section{Acknowledgments}

The authors would like to thank Dr. Alfredo Cuzzocrea for his thoughtful comments on this paper.

\bibliographystyle{abbrv}
\bibliography{bibdatax} 

\end{document}